\documentclass[aps,prl,twocolumn,superscriptaddress]{revtex4}

\usepackage{graphicx,amsmath}
\DeclareGraphicsExtensions{.eps,.pdf}

\begin{document}

\title{Fluctuations and Correlations of Emission from Random Lasers}

\author{Jason W. Merrill}
\affiliation{Department of Physics, Yale University, New Haven, CT 06520}
\author{Hui Cao}
\affiliation{Departments of Applied Physics and Physics, Yale University, New Haven, CT 06520}
\author{Eric R. Dufresne}
\email[]{eric.dufresne@yale.edu}
\affiliation{Departments of Mechanical Engineering and Materials Science, Chemical and Environmental Engineering, Physics, and Cell Biology, Yale University, New Haven, CT 06520}

\date{\today}

\begin{abstract}

When light travels through strongly scattering media with optical gain, the synergy between diffusive transport and stimulated emission can lead to lasing action.
Below the threshold pump power, the emission spectrum is smooth and consistent from shot-to-shot.
Above the lasing threshold, the spectrum of emitted light becomes spiky and shows strong fluctuations from shot-to-shot.
Recent experiments have reported that emitted intensity resembles a power-law distribution (\emph{i.e.} L\'{e}vy statistics).
Recent theories have described the emergence of L\'{e}vy statistics as an intrinsic property of lasing in random media.
To separate intrinsic intensity fluctuations from the motion of scatterers, we compare the statistics of samples with stationary or freely-diffusing scatterers.
Consistent with previous reports,  we observe L\'{e}vy-like statistics when intensity data are pooled across an ensemble of scatterer configurations. 
For fixed scatterers, we find exponential intensity distributions whose mean intensities vary widely across wavelengths. 
L\'{e}vy-like statistics re-emerges when data are combined across many lasing modes.
Additionally, we find strong correlations of lasing peak intensities across wavelengths.
A simple mean-field statistical model captures the observed one- and two-point statistics, where correlations in emission intensity arise from competition among all lasing modes for limited gain.

\end{abstract}


\maketitle

Mesoscopic transport and Anderson localization have been widely studied for electrons, photons and cold atoms. Unlike electrons and atoms, photons may be multiplied via stimulated emission \cite{lagendijk2009fifty}. 
The synergy of coherent amplification and multiple scattering leads to fascinating phenomena such as lasing in random media \cite{cao2003lasing, wiersma2008physics, noginov2005solid}. 
From a basic physics point of view, random lasers are complex, open, nonlinear systems that bridge various fields such as mesoscopic physics, nonlinear dynamics, laser physics, and quantum optics. 
Random lasers exhibit unique characteristics, such as tunable/low spatial coherence \cite{noginov1999, redding2011} and spectral fingerprinting of the random structure \cite{cao2000microlaser}, which points to a wide range of applications in high-speed full-field imaging \cite{redding2012}, optical tagging \cite{wiersma2000laser, lawandy1994}, and cancerous tissue mapping \cite{polson2004random, polson2010}.

Over the past two decades there have been extensive experimental and theoretical studies on random lasers \cite{cao2005review}. 
One fundamental difference from conventional lasers is strong intrinsic fluctuations in the number of lasing modes,  lasing frequencies, and  emission intensities \cite{van2006intrinsic, mujumdar2007chaotic, wu2007statistics, lepri2007statistical, wu2008statistical, zaitsev2009statistical, uppu2010statistical, uppu2012identification, zhu2012, uppu2013dependence, lepri2013, ignesti2013experimental, UppuPRL15}. 
Such fluctuations are attributed to the existence of numerous modes with similar lasing thresholds and strong mode interactions via the gain material. 
One  interesting observation is that the distribution of emission intensities can exhibit features of L\'{e}vy statistics, including a power-law decay. 
The emergence of L\'{e}vy statistics is predicted by analytic and numerical models that account for the statistics of random walks but ignore the interference effect \cite{lepri2007statistical, uppu2013dependence, lepri2013, ignesti2013experimental, UppuPRL15}.
However, most experiments on statistics of random-laser emission were conducted with mobile scatterers, with  random configuration changes from one pump pulse (shot) to the next   \cite{wu2007statistics, wu2008statistical, uppu2010statistical, uppu2012identification, ignesti2013experimental}.
In general, the strong fluctuation of emission spectra from shot-to-shot can arise from (i) amplification of noise from spontaneous emission or fluctuations of the pump, (ii) changes in the positions of mobile scatterers. 
The separate contributions from  different sources to the random laser statistics are not known.   

In this Letter, we separate the contributions of scatterer motion from intrinsic optical fluctuations by comparing the intensity statistics of random laser systems with identical optical properties but different scatterer motion. 
The data from samples with stationary scatterers suggest that individual lasing modes have an exponential distribution of energies while L\'{e}vy statistics emerges when data are pooled across spectrally heterogeneous modes.  
Additionally, we find strong intensity correlations between lasing modes and provide an empirical model that captures the observed one- and two-point statistics.

We consider emission from  colloidal scatterers dispersed in dye-doped liquids. 
Polystyrene scatterers have a radius of  125 nm and are suspended in aqueous media at a number density of about 0.25 $\mu$m$^{-3}$ (0.2 \% volume fraction). 
Using Mie scattering theory, we calculate the scattering mean free path of these samples to be about 410 $\mu$m.
The scatterers are suspended in an aqueous solution of 4.5 mM sulfarhodamine 640 and lauramine oxide, a surfactant used to enhance the solubility of the dye (Ammonyx LO diluted 4:1 with water).  
 
Samples are optically pumped with 30 ps pulses of 532 nm light at a repetition rate of 10 Hz from a frequency-doubled pulsed Nd:YAG laser (Continuum Leopard). 
The pump light is focused by a 2.5 cm focal length lens to a $\approx 10 \mu$m spot. 
The axial position of the cuvette is chosen to maximize the contrast between laser emission peaks and smooth background, which occurs when the excitation pulse is focused a few hundred microns into the sample.
The emitted light in the backward direction is collected by the same lens, passing through a long-pass filter (to remove the scattered pump light) and then directed to a grating spectrometer (Acton 300i).
The exit port of the spectrometer is connected to a CCD array detector, which is synchronized with the pump laser to record the emission spectra of individual pulses. 
When the pump pulse energy exceeds a threshold, discrete lasing peaks appear in the emission spectrum. 
To ensure that the lasing peaks are well-resolved above the amplified spontaneous emission (ASE) background, but also to minimize bleaching of the dye, the incident pump pulse energy is set to 400 nJ, which is roughly twice the lasing threshold.
For each sample we record single-shot emission spectra over 500 pulses at a fixed pump level. 

\begin{figure*} 
\begin{center}
\includegraphics[width=2.\columnwidth]{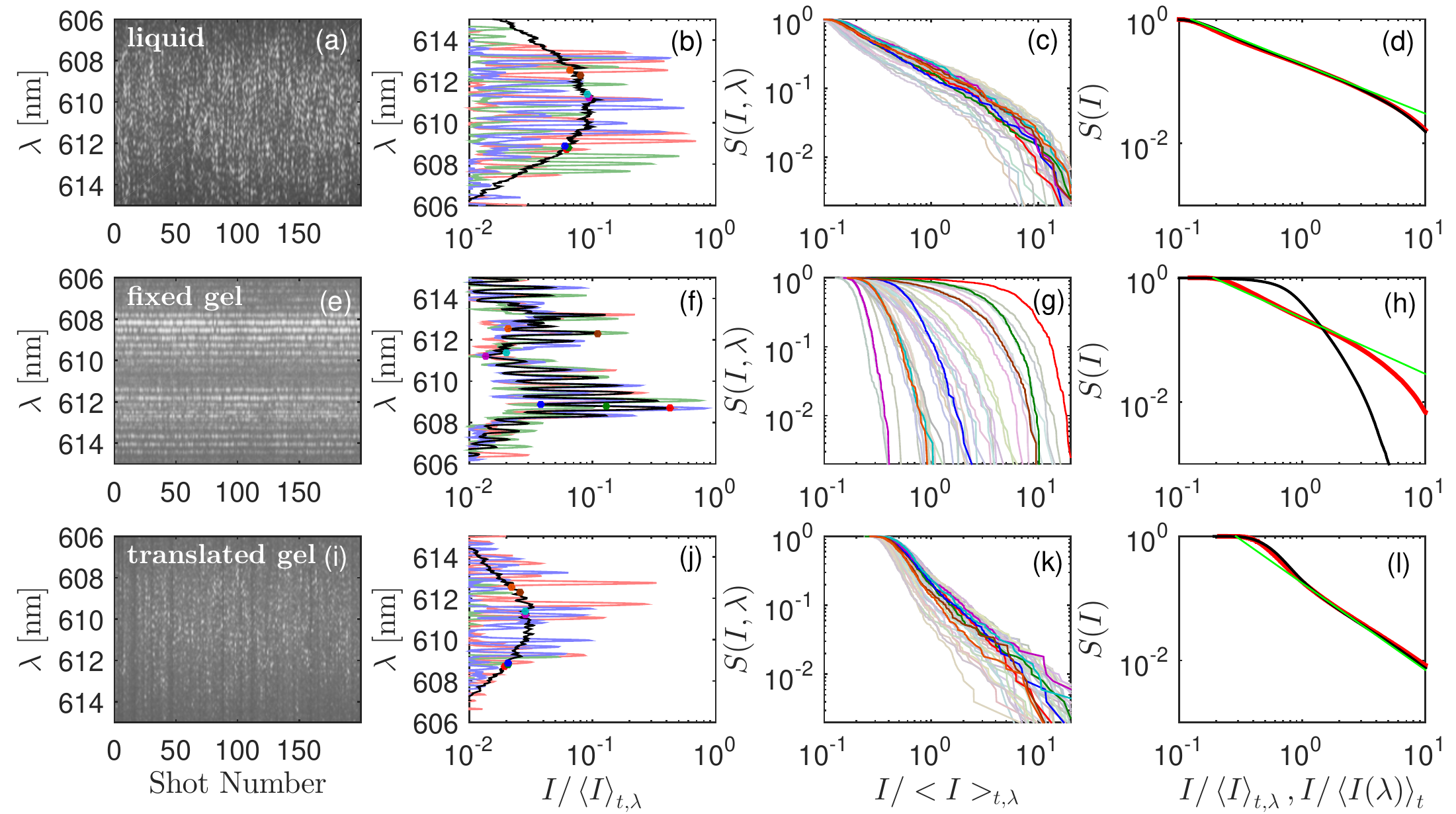}
\end{center}
\caption{\label{fig:raw} \emph{Statistics of Random Lasers with Different Scatterer Motion.}  Data from a liquid sample, stationary gel sample and 100 $\mu$m/s translating gel sample are shown in the first row (a-d), second row (e-g) and third row (i-l), respectively.  The first column (a,e,i) shows emission spectra for 200 consecutive shots.  Second column (b,f,j) shows the mean energy over 500 shots at each wavelength (thick black line), and the intensities collected in 3 individual shots (red, green, and blue lines). 
 The third column (c,g,k) shows distributions of laser emission intensity for individual wavelengths  between 608 and 614 nm. 
The colored distributions correspond to wavelengths highlighted by the same colored dots in the second column.  The fourth column (d,h,l) shows the intensity distribution when data are pooled across wavelenths.  The red curve shows the distribution of the intensities pooled across shots and wavelengths, normalized by the mean over all shots and wavelengths $\left<I\right>_{t, \lambda}$ . The black curve shows the distribution of intensities that are first normalized by the average intensity at each wavelength $\left<I(\lambda)\right>_t$ and then pooled across wavelengths.   
 }
\end{figure*}

Results for scatterers immersed in the liquid gain medium are shown in the first row of Fig. \ref{fig:raw}.
Raw emission spectra, $I(\lambda,t)$, from 200 consecutive shots are shown in Fig. \ref{fig:raw}{a}.  
The emission spectrum of each shot is composed of about twenty discrete peaks, and the peak frequencies change completely from shot-to-shot. 
Thus, each peak appears as an isolated bright spot on the 2D plot of emission spectra across shots in Fig. \ref{fig:raw}(a).
While emission peaks dominate single-shot spectra,  the emission spectrum averaged over 500 shots $\left<I(\lambda)\right>_t$ is smooth, as shown in   Fig. \ref{fig:raw}(b).
Each wavelength shows large fluctuations in the intensity of emission, and we see no correlation in the location or height of the peaks from shot-to-shot.  
We quantity the intensity fluctuations at each wavelength using the survival function, $S(I,\lambda)$, in Fig. \ref{fig:raw}(c).
The survival function, $S(I,\lambda)$, gives the fraction of shots with emission intensity at wavelength $\lambda$ above the intensity $I$. 
Compared to density histograms, this method of quantifying fluctuations has better fidelity in the tail because there is no binning and every intensity value is plotted. 
Note that exponential and power-law probability distributions give exponential and power-law survival functions.
Here, the survival functions fall off with a power-law tail with exponent $\nu \approx -1$ before reaching a cut-off at near 10 times the mean intensity.
This is consistent with previous reports \cite{wu2007statistics, wu2008statistical, uppu2010statistical, uppu2012identification, ignesti2013experimental} of L\'{e}vy statistics of the emission of random lasers.
The fourth panel shows the distribution of intensities pooled across  wavelengths.
For the red curve, $S(I/\left<I\right>_{t, \lambda})$, the intensities are pooled directly across shots and wavelengths, and normalized by the mean over all shots and wavelengths $\left<I \right>_{t, \lambda}$. 
For the black curve, $S(I(\lambda)/\left<I(\lambda)\right>_t)$, the intensities at each wavelength are first normalized by the average intensity of that wavelength $\left<I(\lambda)\right>_t$ before pooling across wavelengths.
These two distributions are essentially identical, suggesting that the  distribution of intensities is similar across the emission spectrum.

In the above experiments, the structure of the scatterers changes completely between successive shots.
In this time inverval (100 ms), the scatterers diffuse a distance comparable to their separation, about 1 $\mu$m.
To remove this source of fluctuations, we arrest scatterer motion by gelling the liquid. 
A gelatinous gain medium allows for the free diffusion of water and dye, but blocks the translation of larger particles.  
We use a gel which is 92\% liquid and 8\% polyacrylamide with a Young's modulus of about 1KPa \cite{wang1998}.  
Under these conditions, the scattering strength and the amount of gain material in the gelatinous sample are nearly identical to those of the fluid sample, but now the scatterers are fixed with residual fluctuations of about 5 nm.

The fluctuations of emission from the gel sample are distinct from the liquid sample, as shown in the second row of Fig. \ref{fig:raw}.
The discrete lasing peaks do not change location from shot-to-shot, Fig. \ref{fig:raw}(e), and do not average into a smooth spectrum upon averaging across shots, Fig. \ref{fig:raw}(f).  
In each shot, the emission is randomly distributed among discrete peaks, and there is no correlation in single-peak height fluctuations from shot-to-shot.
The single-wavelength intensity distributions do not show the slow decay characteristic of  L\'{e}vy statistics that was observed for the liquid sample, Fig. \ref{fig:raw}(g).
While the statistical distributions of emission intensities at individual wavelengths in the liquid sample are very similar, the median intensities vary widely across wavelength in the gel sample.  
When the emission data are pooled across wavelengths, Fig. \ref{fig:raw}(h), a power-law distribution re-emerges with exponent $\nu \approx -1$.
However, when the intensities at each wavelength are normalized to the average of that wavelength $\left<I(\lambda)\right>_t$ and then pooled across wavelengths, the distribution becomes similar to that of a single wavelength and decays much more rapidly than the power-law.

Why do the liquid and gel gain media show different emission statistics?
The gain materials and pump conditions are the same in both cases. 
The two samples have identical scatterers embedded in media with nearly identical optical properties. 
Therefore, we do not expect the statistical differences between these samples to reflect differences in  lasing action.
Instead, the dominant difference between these two samples lies in the dynamics of the scatterers. 
These dynamical differences have no significant impact on lasing action within a single shot because samples have essentially stationary scatterers over the duration of a single pump pulse:
over this time interval (30 ps), the particles diffuse about 0.1 \AA.
Therefore, the remaining difference lies in the ensemble of scatterer configurations sampled by the series of pump pulses.
While the particle configurations are uncorrelated from shot-to-shot in the liquid gain medium, they are essentially identical across all shots in the gelatinous gain medium.  
This explains our observation that the location of the peaks in the emission spectra of the gel are consistent from shot-to-shot.
To confirm that the differences in the statistics of the liquid and gelatinous media are due to particle motion, we acquired random lasing spectra from a gel sample while translating the pump spot on the sample at a speed of 100 $\mu$m/s.  
At this speed of translation, there is little overlap in the pump volume from shot-to-shot, so the sampled scatterer configurations are independent.
As shown in the third row of Fig. \ref{fig:raw},  the emission spectrum of translated gel recapitulates all of the features of the  liquid sample.

\begin{figure*} 
\includegraphics[width=2.\columnwidth]{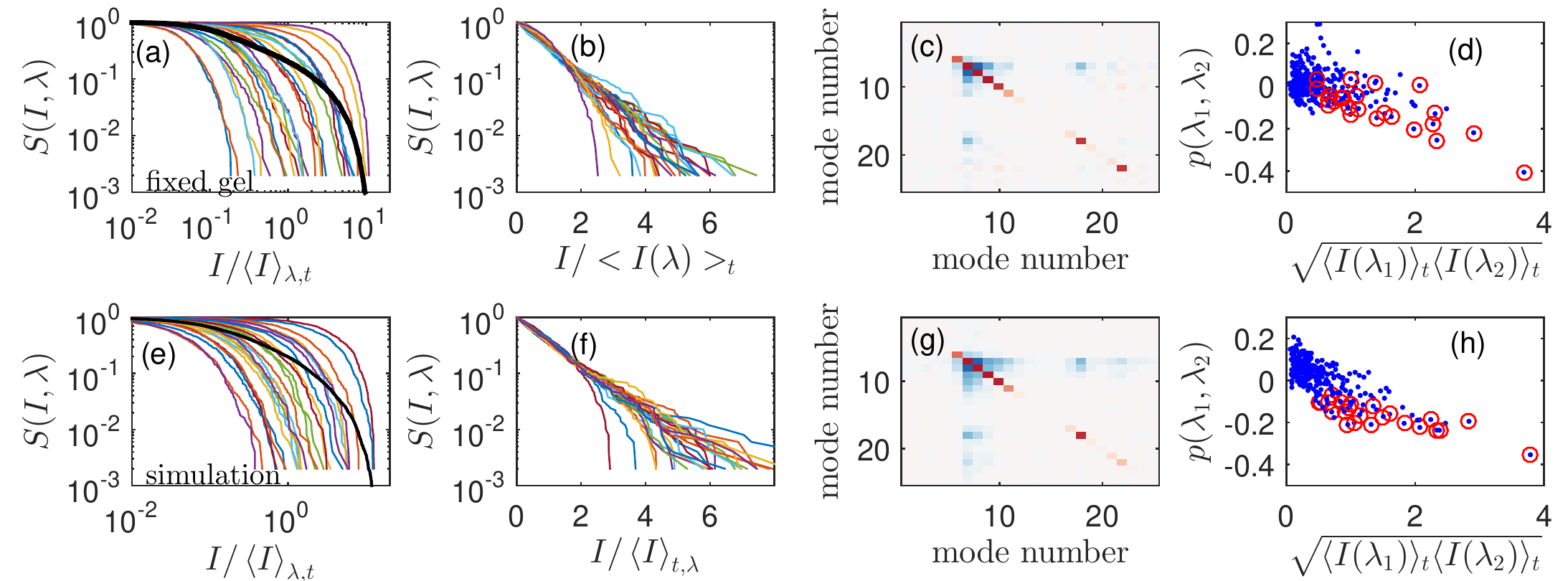}
\caption{\label{fig:track} Statistics of random lasing modes in a gel sample with immobile scatterers (first row, a-d) and a simple mean-field simulation (second row, e-h).  First column (a,e): colored curves show the survival function of intensities for all 25 lasing modes on a log-log plot.  The thick black curves show the survival functions of intensity pooled across all of the modes.  Second column (b,f) shows survival functions of intensity normalized to the average of each mode on a semi-log plot.   Third column (c,g) shows the correlations (covariance) across lasing modes.  Red (blue) indicates positive (negative) correlations. 
The diagonal is the variance of each peak, and the off-diagonal shows correlations across peaks.  
Fourth column (d,g): scatter plot of Pearson correlations between lasing modes (excluding self-correlations).  The data points with red circles indicate correlations with the brightest lasing mode.}
\end{figure*}

Having experimentally removed the effects of scatterer motion, we quantify the shot-to-shot fluctuations of indiviudal lasing mode intensities.
Locating peaks in the mean emission spectrum of the stationary gel sample (Fig. \ref{fig:raw} second row),  we identify 25 distinct lasing modes.
The survival functions, $S(I,\lambda)$, for each mode are shown as colored lines in Fig. \ref{fig:track}a.
When we normalize the intensity of each mode by its mean and plot the resulting survival function, we find that all but the brightest modes have an exponential distribution of intensities, as shown in Fig. \ref{fig:track}b.
Therefore, at the level of a single lasing mode, we find no signature of L\'{e}vy statistics.
However, the mean intensity varies widely from mode to mode, so that if we pool the data across all modes before computing the survival function, the statistical distribution of emission intensity displays a slower-than-exponential decay, as shown by the thick black line in Fig. \ref{fig:track}a.

We also find that the fluctutations of emission intensity   are coupled across lasing modes.
We quantify  cross-correlations in the lasing mode intensities using the covariance: \begin{equation}
  c_2(\lambda_1, \lambda_2) = \langle I(\lambda_1) I(\lambda_2)\rangle_t - \langle I(\lambda_1)\rangle_t \langle I(\lambda_2)\rangle_t.
\end{equation}
 The lasing modes can be significantly correlated, even when they  are well-separated across wavelengths, as seen in Fig. \ref{fig:track}c.
Since the  intensity distribution of each mode is nearly exponential, $c_2(\lambda, \lambda)  \approx \langle I(\lambda) \rangle_t^2$. 
Thus, the brightest modes have the highest variances, which are shown along the diagonal in Fig. \ref{fig:track}c.
 These bright modes are anti-correlated with other modes, indicated by the off-diagonal elements. 
To scale out the trivial dependence of the cross-correlations on the variance of each mode, we calculate the Pearson correlation between each pair of modes, $ p(\lambda_1,\lambda_2)= c_2(\lambda_1, \lambda_2) /\sqrt{c_2(\lambda_1,\lambda_1) c_2(\lambda_2,\lambda_2)}$.
As shown in Fig. \ref{fig:track}d,  the strength of the Pearson correlation between modes scales with the geometric mean of their intensities.
The strongest cross-correlations are those involving the brightest mode, indicated by red circles in Fig. \ref{fig:track}d.

The brightest lasing modes can saturate the optical gain and suppress lasing in other modes, leading to the anti-correlations of intensities. 
Moreover, the intensity distribution of the brightest modes decay faster than exponentially, as a result of gain depletion.  
To construct an empirical model to capture these features, we take a mean-field approach. 
For each pump pulse, we assign an random initial intensity to the $n$-th mode, \emph{e.g.}, $I_n^{(i)}$.
We assume that the statistical distribution of the initial intensity of each mode is exponential, $\rho(I_n^{(i)})=\langle I_n^{(i)} \rangle  \exp[{-I_n^{(i)}/ \langle I_n^{(i)} \rangle}]$.
Since the pump pulse energy is identical for each shot, the total emission intensity should be constant.  
We determine the observed intensities by renormalizing the initial intensities of individual modes by the sum of the initial intensities across the modes:  $I_n = I_n^{(i)}/ \sum_m I_m^{(i)}$.
This normalization enforces global energy conservation, thus taking into  account the mode competition for a limited gain.
To compare this simple model to our data, we assign the experimental value of the observed mean intensity for each lasing mode $\langle I_n \rangle_t$ to $\langle I_n^{(i)} \rangle$.
The results for this simple model with 500 simulated shots are shown in the second row of Fig. \ref{fig:track}. 
The model captures the essential features of the one-point statistics: most of the modes decay exponentially, while the brightest ones decay more quickly due to gain saturation. 
Importantly, the pooled-intensity statistics across the modes show a slower-than-exponential decay. 
The simulation also captures the correlated intensity fluctuations between lasing modes, with the strength of the anti-correlation increasing with the geometric mean of the intensity of the two modes.

This simple model faithfully reproduces the observed one- and two-point statistics of random lasers with fixed scatterers.
Since a mean-field enforcement of energy conservation effectively simulates gain competition and saturation, we expect our experiment has a strong spatial overlap of the lasing modes.
However, to specify the mechanism of mode-competition, the self- and cross-saturation coefficients of lasing modes will need to be calculated by the first-principle scattering and lasing theory \cite{tureci2008strong, tureci2009ab}, taking into account their spatial and spectral overlaps \cite{cao2005review}. 
A rigorous physical model remains to be developed to address the exponential distribution of the inherent intensities and the wide variation of mean intensities across modes.

More generally, this work empirically separates contributions to random laser fluctuations from scatterer motion and inherent optical fluctuations such as spontaneous emission.  
For the parameters probed in this experiment, we find that the inherent intensity distribution of each random laser mode is exponential, and power-law distributions emerge only when scatterers move or data are pooled across modes with widely varying average intensities.
These results provide physical insight into random laser statistics, and will stimulate further studies.
Our results also indicate that the statistics of random lasing spectra reflect the microscopic motion of underlying scatterers.  
This suggests that random laser statistics may be able to quantify the motion of microscopic particles in dense suspensions where conventional dynamic light scattering methods are limited.

We acknowledge helpful discussions with Heeso Noh, Brandon Redding, Yindong Chong and A. Douglas Stone and financial support from the donors of the ACS Petroleum Research Fund (50872-ND6), the National Science Foundation (CBET-1236086) and the Office of Naval Research (MURI SP0001135605). 


\end{document}